\documentstyle[aps,pre]{revtex}

\newcommand{\be}{\begin{equation}}
\newcommand{\ee}{\end{equation}}
\newcommand{\BE}{\begin{eqnarray}}
\newcommand{\EE}{\end{eqnarray}}
\newcommand{\barr}{\begin{array}}
\newcommand{\earr}{\end{array}}

\begin{document}

%\title{An approach to the Multiple Scale Expansion technique in 
%nonlinear discrete  Klein-Gordon vector lattices }
%\title{How to find small amplitude envelope soliton solutions
%in nonlinear discrete  Klein-Gordon vector lattices }
\title{Vector Nonlinear Klein-Gordon Lattices:
General Derivation of Small Amplitude Envelope Soliton Solutions} 
\author{Simona Cocco$^{1,3}$, Maria Barbi$^{2,3}$ and Michel Peyrard$^3$}

\address {$^1$  Dipartimento di  Scienze Biochimiche
Universit\`a di Roma "La Sapienza"
P.le A. Moro, 5 - 00185 Roma, Italy}

\address{$^2$ Istituto Nazionale di Fisica Nucleare, Dipartimento di
Fisica, Universit\`a degli Studi di Firenze, Largo E. Fermi, 2 - 50125
Firenze, Italy}

\address{$^3$ Laboratoire de Physique, CNRS URA 1325, Ecole Normale
Sup\'erieure de Lyon, 46 All\'ee d'Italie, 69364 Lyon Cedex 07,
France.}

\date{\today}

\maketitle

\begin{abstract}

Group velocity and group velocity dispersion for a wave packet in 
vectorial  discrete Klein-Gordon models are obtained 
by an  expansion, based on  perturbation theory,
 of the linear system giving the dispersion relation and the normal modes.
 We show how to map this expansion on the 
Multiple Scale Expansion in the real space and  how to find Non Linear 
Schr\"odinger small amplitude solutions  when a nonlinear
one site potential balances  the group velocity  dispersion effect.

\end{abstract}

%%%%%%%%%%%%%%%%%%%%%%%%%%%%%%%%%%%%%%%%%%%%%%%%%%%%%%%%%%%%%%%%%%%%%%%%%%%

\section{Introduction.}

One of the most popular approaches used to determine the small amplitude
 envelope
soliton solutions in non linear models is the well-known
 Multiple Scale  
Expansion (MSE) technique \cite{rem}. 
This  technique amounts to expanding the equations of motion on
different time and space scales 
 looking  for wave-packet-like solutions; a wave packet is 
a superposition of plane waves whose frequencies and wave vectors lie 
in a narrow band, and it can be conveniently described by a plane wave 
with an amplitude that varies slowly in space and time.
Increasing  progressively the time and space
scales one  determines in a first step
the carrier wave  as  a phonon  mode of the linearized  system,
then deduces the partial differential equation that identifies the envelope
phase velocity with the wave packet group velocity, and finally derives
the NLS equation for the envelope whose  diffusion coefficient is in fact
the  wave packet  group velocity  dispersion. 
An alternative method, commonly used in optics,
consists in expanding the dispersion relations with
respect to the carrier frequency and then in  building at each order of this expansion
an operator that acts on the envelope function
\cite{opt}.
 MSE has been successfully applied to various non linear systems
with scalar fields and  the corresponding method adopted in optics 
has permitted the study of optical solitons in fiber.
In this latter case  one  deals with eletric field components which are
not coupled
at the linear order  of the Maxwell equations.
 However, many non linear models of interest
involve vectorial fields with coupled components at the linear
order that give rise to  dispersion relations with more than one branch:
 classical examples are given by
multi-atomic lattices, or by lattices in which the mass at
each site can move in a multidimensional space.  

In this work, we show how to find small amplitude envelope soliton solutions 
in such vectorial lattices problems.
 The main difficulty with respect
to the scalar case is to determine the relative amplitudes of the
different components of the field.
We will perform a  perturbative expansion, around one
wavenumber, of the linear system that gives the dispersion relations
and the linear eigenmodes; then
we will   introduce an operator 
formalism analogous to that used in optics to obtain, from this
expansion and from the nonlinear terms,
    the MSE 
equations, up to the NLS one. 

An application of the method presented in this work
can be found in a forthcoming paper \cite{noi}, where
the envelope
soliton solutions of an helicoidal DNA model 
described by a radial and an angular degree of freedom for each site
are derived.

%%%%%%%%%%%%%%%%%%%%%%%%%%%%%%%%%%%%%%%%%%%%%%%%%%%%%%%%%%%%%%%%%%%%%

\section {Wave-packet in linear vectorial lattices}

The NLS equation is obtained when a weak dispersion is balanced by a
weak nonlinearity. In order to characterize the dispersion, let us
first restrict our attention to the linear part of the system of interest.
We start with a one dimensional vectorial linear lattice model 
 given by the equations of motion:

\be
\label{emoto}
\frac{\partial^2 E(n,\alpha,t)}{\partial t^2 }=
- \sum_{n',\alpha'} J (n-n',\alpha,\alpha') E(n',\alpha',t) 
\ee
where  $n,n'$ are the site indices, $\alpha,\alpha'$ are the 
indices that label the components of the vectorial field
$ E(n,\alpha,t)$, and 
$J (n-n',\alpha,\alpha')$ are the force constants depending on 
$ n-n'$ for translationally invariant systems.

Looking for  plane wave solutions of the form  
\be 
A \vec{V}_l(q) e^{i(q n-\omega_l(q)t)} + c.c.
\ee  
where  $A$ is the wave amplitude,  the equation of motion is mapped  to the 
operator equation in the wave numbers space:      
\be
\label{d}
(\hat{J}(q)- \omega^{2}_{l}(q))\vec{V}_{l}(q) =0 
\ee
where $\hat{J}(q)$ is the Fourier transform of the matrix $\hat{J}(n-n')$.
The index $l$ runs from 1 to the number of components of the vectorial
field $E(n,\alpha,t)$; the eigenvalues functions  $\omega^{2}_{l}(q)$ 
give the branches of the dispersion relation;
the normal modes $\vec{V}_l(q)$ are the orthonormalized eigenvectors of
the matrix $\hat{J}(q)- \omega^{2}_{l}(q)  $.  

In order to investigate the dispersion, we now consider a wave-packet-like
solution, {\em i.e.} a superposition of plane waves with wave numbers 
in a small interval:
\be 
\label{wpuno} 
\vec{E}_l(n,t)=\int_{q_0 - \Delta q}^{q_0 + \Delta q} 
 A(q) \, \vec{V}_{l}(q) \,
 e^{i(q n-\omega_l(q)t)} \; dq +c.c.
\ee
 
For each $q$ contributing to the wave packet the system of equations
 (\ref{d}) must
be fulfilled. The weakly dispersive case 
is obtained by considering only small deviations of $q$ with
respect to the wavector $q_0$ corresponding to the center of the
wavepacket. To measure this deviation, the wavevector $q$ is written $
q = q_0 + \epsilon q_1$, where $\epsilon \ll 1$.

 Eq.~(\ref{d}) is
solved, $\forall q_1$ in the integration
range, by a perturbative technique.

The operator $\hat{J}(q_0+\epsilon q_1)$ can be expanded in Taylor series as 
$\hat{J}(q_0)+\epsilon \hat{J}'(q_0)q_1+\epsilon^2
\hat{J}''(q_0)q_{1}^{2}/2 + \ldots$. The quantities 
$\epsilon \hat{J'}(q_0)q_1$, $\epsilon^2 \hat{J''}(q_0)q_1^2/2$ 
are small perturbations   
with respect to unperturbed  operator   $\hat{J}(q_0)$, 
whose eigenvalues are $\omega^2_l=\omega^2_l(q_0)$ and whose 
eigenvectors $\vec{V}_{l}=\vec{V}_{l}(q_0)$ constitute a complete basis.
  
According to standard perturbation theory \cite{coh} we 
 write  the expansions of the eigenvectors and eigenvalues, 
 \BE
\label{v}
\vec{V}_{l}(q_0+\epsilon q_1) &=&\vec{V}_{l}+ \epsilon \vec{V}^{(1)}_{l}q_1
+ \epsilon^2 \vec{V}^{(2)}_{l}q_1^2/2 +\ldots \\  
\label{omega}
\omega_l(q_0+\epsilon q_1) &=&\omega_l+\epsilon 
\omega^{(1)}_l q_1 +
\epsilon^2  \omega^{(2)}_l  q_1^2/2+ \ldots
\EE

Equation (\ref{d}) has to be solved at each expansion order:
\BE
\label{o1} \text{at order $\epsilon^0$ : }\quad
\hat{J}\vec{V}_l &=& \omega^2_l \vec{V}_l \\ 
\label{o2}
\text{at order $\epsilon^1$ : }\quad
(\hat{J} \vec{V}^{(1)}_l+\hat{J}' \vec{V}_l)\;q_1 &=& 
(2\omega_l \omega^{(1)}_l \vec{V}_l+ \omega^2_l \vec{V}^{(1)}_l)\;q_1 \\ 
\label{o3}
\text{at order $\epsilon^2$ : }\quad
(\hat{J}' \vec{V}^{(1)}_l + \frac{1}{2} \hat{J}'' \vec{V}_l+ 
\frac{1}{2} \hat{J} \vec{V}^{(2)}_l)\;q_1^2 &=&
{(\omega^{(1)}}^2_l \vec{V}_l +\omega_l \omega^{(2)}_l\vec{V}_l
+2\omega_l \omega^{(1)}_l \vec{V}^{(1)}_l+
\frac{\omega^2_l}{2} \vec{V}^{(2)}_l)\;q_1^2 \,.
\EE

At order $\epsilon^0$, 
one solves the unperturbed problem determining $\vec{V}_l$
and $\omega_l$. At order $\epsilon$, one determines $\vec{V}^{(1)}_l$,
$\omega^{(1)}_l$: imposing to $\vec{V}^{(1)}_l$ to be orthogonal to $\vec{V}_l$ to 
guarantee the normalization of $\vec{V}_l(q_0+\epsilon q_1)$  the scalar 
 product of (\ref{o2}) with  $\vec{V}^*_m \;\; (m\neq l)$ gives
\BE
\label{vp}
\vec{V}^{(1)}_l= \sum_{m\neq l} \alpha_m \vec{V}_m \\
\label{am}
\alpha_m= \frac{ {\vec{V}_m}^* \hat{J}' \vec{V}_l}{ \omega^2_l-\omega^2_m} 
\;\;\;\;\;\;\; m \neq l
\EE
and that with  $\vec{V}^*_l$ gives
\be
\label{op}
\omega^{(1)}_l=  \frac{ {\vec{V}_l}^* \hat{J}' \vec{V}_l}{2 \omega_l}\,.
\ee
At order $\epsilon^2$, one determines  $\omega^{(2)}_l$  by 
multiplying (\ref{o3})  by  $\vec{V}^*_l$
\be
\label{os}
\omega^{(2)}_l= \frac{1}{\omega_l} \left({\vec{V}_l}^* \frac{\hat{J}''}{2} \vec{V}_l
-{\omega^{(1)}}^2_l+ 
 \sum_{m\neq l} \frac{| {\vec{V}_m}^* \hat{J}' \vec{V}_l|^2} {
\omega^2_l-\omega^2_m} \right)\,.
\ee
We  assume, for sake of simplicity,  that the eigenmodes of J
 (see (\ref{d})) 
are not degenerate, but the generalization of the degenerate
 case is straightforward with the standard perturbation theory.

The phase of each component of (\ref{wpuno}) can be expanded  around
the central wave number  $q_0$,  up to second order in $\epsilon q_1=q-q_0$
using the values  of $\omega^{(1)}_l$ and $\omega^{(2)}_l$ determined above
\be 
\label{wp}
 \vec{E}_l(n,t)= e^{i(q_0 n-\omega_l(q_0)t)}
\epsilon \int_{-\Delta q/\epsilon}^{+\Delta q/\epsilon} 
A(q_0+ \epsilon q_1) \vec{V}_{l}(q_0+ \epsilon q_1) 
\;
\exp{\left\{i \epsilon q_1(n-\omega^{(1)}_l(q_0)t)
-i \epsilon^2 \frac{q_1^2}{2}\omega^{(2)}_l(q_0)t\right\}}
\; dq_1+ c.c. 
\ee

Under this form, $\vec{E}_l(n,t)$ appears as a plane wave, henceforth
called the carrier wave, with an
amplitude that depends on space and time and which corresponds to 
the integral of equation (\ref{wp}), 
$\vec{E}_l(n,t) =  \vec{F}(n,t) \; \exp[i(q_0 n-\omega_l(q_0)t)] +c.c.$. 
The fact that $\omega^{(1)}_l(q_0)$, 
$\omega^{(2)}_l(q_0)$ obey  relations (\ref{op}) and (\ref{os}) ensures
that this wave packet is a solution of the original equation
(\ref{emoto}), up to the order of the various expansions. 

In order to
extend the study to the nonlinear case it is useful to express these
conditions under the form of an equation in the space time coordinates
for the amplitude. Let us introduce the quantity
\be 
\label{ap}
 A(n,t)= \int_{-\Delta q/\epsilon}^{+\Delta q/\epsilon} A(q_0+ \epsilon q_1)
\;
\exp{\left\{i \epsilon q_1(n-\omega^{(1)}_l(q_0)t)
-i \epsilon^2 \frac{q_1^2}{2}\omega^{(2)}_l(q_0)t\right\}}
\; dq_1 
\ee
Equation (\ref{ap}) shows that $A(n,t)$  slowly varies in space and
time. In the spirit of the multiple scale expansion, it is natural to
introduce the slow variables $x_1 = \epsilon n$, $t_1 = \epsilon t$
and $t_2 = \epsilon^2 t$ so that $A(n,t)$ can be written as
\be
\label{ap1}
 A(n,t) = A(x_1,t_1,t_2) =
 \int_{-\Delta q/\epsilon}^{+\Delta q/\epsilon} A(q_0+ \epsilon q_1)
\;
\exp{\left\{i q_1( x_1-\omega^{(1)}_l(q_0)t_1)
-i \frac{q_1^2}{2}\omega^{(2)}_l(q_0)t_2\right\}}
\; dq_1 
\ee
or
\be
\label{ap2}
 A(x_1,t_1,t_2) = A(s_1,t_2) = 
\int_{-\Delta q/\epsilon}^{+\Delta q/\epsilon} A(q_0+ \epsilon q_1)
\;
\exp{\left\{i q_1 s_1
-i \frac{q_1^2}{2}\omega^{(2)}_l(q_0)t_2\right\}}
\; dq_1  
\ee
with the introduction of the variable $s_1 = x_1 - \omega^{(1)}_l(q_0)t_1$
 to switch to the frame moving at the group velocity of
the carrier wave.

Using the relation $(\partial A / \partial x_1) = (\partial A /
\partial s_1) = i<q_1> \equiv 
\int_{-\Delta q/\epsilon}^{+\Delta q/\epsilon} 
i q_1\; A(q_0+\epsilon q_1)
\;
\exp{\left\{i q_1 s_1
-i \frac{q_1^2}{2}\omega^{(2)}_l(q_0)t_2\right\}}
\; dq_1$  that derives directly from
Eqs.~(\ref{ap1}) and (\ref{ap2}), and the expansion (\ref{v}) of $V(q_0
+ \epsilon q_1)$, the amplitude $\vec{F}$ of the wave can be
expressed as a function of $A(s_1,t_2)$ by the relation
\be
\label{f}
\vec{F}(x_1,t_1,t_2)= \epsilon
(\vec{V}_{l}-i 
\epsilon \vec {V'_l} \frac{\partial }{\partial x_1}) A(x_1,t_1,t_2) 
\; .
\ee

{From} (\ref{ap1}) and (\ref{ap2}), we directly derive
the equations of motions of $A$ as a function of the slow space--time
variables:
\be
\label{wp1}
\left( \frac{\partial A}{\partial t_1}+
 \omega^{(1)}_l\frac{\partial A}{\partial x_1} \right) =0
\ee
and 
\be
\label{ls}
i \frac{\partial A }{\partial t_2}+
 \frac{\omega^{(2)}_l}{2}\frac{\partial^2 A}{\partial s_1^2}=0
\ee
where $\omega^{(1)}_l$ and $\omega^{(2)}_l$ are then the group velocity and
the group velocity dispersion of the wave packet and determine
the peak velocity and the spread out of the envelope function.
Equation (\ref{f}) shows that
$\vec{V'}_{l}$ determines the first order correction to the 
direction of the  vectorial field solution.

%%%%%%%%%%%%%%%%%%%%%%%%%%%%%%%%%%%%%%%%%%%%%%%%%%%%%%%%%%%%%%%%%%%%%%%
\section{Non Linear Vectorial Lattice }

We now consider the full equation of motion, including
nonlinear on-site potential terms.
Extra nonlinear terms depending on
time derivatives may also appear in the case of non Cartesian coordinates systems:
\BE
\label{emnl}
\frac{\partial^2 E(n,\alpha,t)}{\partial t^2 } =
- \sum_{n',\alpha'} J (n-n',\alpha,\alpha') E(n',\alpha',t)+ \nonumber\\
\sum_{d=0}^{2} \sum_{k=0}^{d}\sum_{\alpha'',\alpha'\leq \alpha''} 
c_{d,k}^{\alpha}(\alpha',\alpha'') E^{(k)}(n,\alpha',t)E^{(d-k)}(n,\alpha'',t)
+ \nonumber\\
\sum_{d=0}^{2} \sum_{k=0}^{d}\sum_{j=0}^{k}
\sum_{\alpha''',\alpha''\leq \alpha''', \alpha'\leq \alpha'' } C_{d,k,j}^{\alpha}(\alpha',\alpha'',\alpha''')
E^{(j)}(n,\alpha',t)E^{(k-j)}(n,\alpha'',t)E^{(d-k)}(n,\alpha''',t) 
\EE
where $E^{(j)}(n,\alpha,t)$ indicates the $j-$th time derivative of    
$E(n,\alpha,t)$, and  $c_{d,k}^{\alpha}(\alpha',\alpha'')$,
$C_{d,k,j}^{\alpha}(\alpha',\alpha'',\alpha''')$ are  the quadratic
and cubic nonlinear terms numerical  coefficients. Index $d$ is the total
time derivative order of each term. The terms with $d=0$ are the
nonlinear potential force terms; the others derive from the kinetic 
energy in the case of non Cartesian coordinates so that $d \leq 2$.

The quadratic terms in (\ref{emnl}) give rise to second  harmonic and
constant terms that, if  we
look  for small amplitude  solution,  have to be included  as
additional smaller corrections:
\BE
\label{wpc}
\vec{E}_l(n,t) &=& \epsilon \, e^{i(q_0n_0-\omega_l t_0)}\,  
 (\vec{V}_{l}-i \epsilon \vec {V^{(1)}_l} \frac{\partial }{\partial x_1})
 A(x_1,t_1,t_2) \nonumber \\
&+& \epsilon^2 \,
 e^{2i(q_0n_0-\omega_l t_0)}\, 
\vec\gamma_l \, A^2(x_1,t_1,t_2)
+\epsilon^2 \,  \vec\mu_l \, |A(x_1,t_1,t_2)|^2 + O(\epsilon^3) \,.
\EE
We are interested in situations where dispersion can balance
nonlinearity, and therefore have to  be measured by the same scaling
parameter $\epsilon$. While the overall $\epsilon$ factor was not
important in the linear case, it must be explicitly included in the
nonlinear case.

We solve the equation of motion (\ref{emnl}) on the three 
characteristic magnitude scales of the wave
packet. At order $\epsilon$ we get 
$\omega_l$ ,$\vec{V}_{l}$ from equation (\ref{o1}).

At order $\epsilon^2$  we get for  the wave packet term 
expressed in the form (\ref{wpuno})  
  the 
system of equations (\ref{o2}) for each $q_1$ in the integration
range. After integration
on the envelope distribution this gives rise to the equation in the
anti-transformed  Fourier space
\be
\label{oe}
 \left( 2\omega_l  \vec{V}_l \frac{\partial }{\partial t_1} + 
(\hat{J} - \omega^2_l) \vec{V}^{(1)}_l \frac{\partial }{\partial x_1}
+ \hat{J}' \vec{V}_l \frac{\partial }{\partial x_1} \right) 
A(x_1,t_1,t_2)=0 \; .
\ee
In fact from (\ref{ap1}), the average wave numbers deviation is  
 $ <q_1>= -i \partial A / \partial x_1$ and the averaged frequency deviation
is in the same way
 $<\Delta \omega_l>=<\omega_l^{(1)}q_1>= i \partial A / \partial t_1$. 

By scalar product of (\ref{oe}) and $V^*_m$  $ \forall m\neq l$  we obtain  the
components   $\alpha_m$ (\ref{am}) of $\vec{V}^{(1)}_l$ (\ref{vp}) on the   
base $\{\vec{V}_m\}$ and by scalar product of (\ref{oe}) and $V^*_l$ we obtain
the equation  (\ref{wp1}) with $\omega^{(1)}_l$ defined by (\ref{op}).

At the same order of expansion one  determines the  vectors
$\vec{\gamma}_l$, $\vec{\mu}_l$  collecting the  terms of
corresponding order $(\epsilon^{2})$ and phase in the equation of
motion (\ref{emnl}) in which the solution form (\ref{wpc})
has been inserted \cite{nota}. One then obtains
 $\vec{\gamma}_l$ by solving the algebraic system
\be
\label{ga}
(\hat{J}(2q_0)-4 \omega^2_l(q_0))\vec{\gamma}_l= 
\sum_{d=0}^{2}\sum_{k=0}^{d} \sum_{\alpha'',\alpha'\leq \alpha''}
\vec{c}_{d,k}(\alpha',\alpha'') (-i\omega_l)^d V_l(\alpha')V_l(\alpha'')
\ee
where  $\vec{c}_{d,k}(\alpha',\alpha'')$ is the vector of components 
${c}^{\alpha}_{d,k}(\alpha',\alpha'')$, each derivative with respect
to $t_0$  giving a factor  $(-i\omega_l)$.
And $\vec{\mu}_l$ is obtained from the system
\be
\label{mu}
\hat{J}(0)\vec{\mu}_l= 
\sum_{d=0}^{2}\sum_{k=0}^{d} \sum_{\alpha'',\alpha'\leq \alpha''}
\vec{c}_{d,k}(\alpha',\alpha'')[(i\omega_l)^k (-i\omega_l)^{d-k}
V^*_l(\alpha')V_l(\alpha'')+(-i\omega_l)^k (i\omega_l)^{d-k}
V^*_l(\alpha'')V_l(\alpha')] \,.
\ee

At order $\epsilon^3$,  from (\ref{emnl}) and (\ref{wpc})
for the  terms in $ e^{i(q_0n_0-\omega_l(q_0)t_0)}$, one obtains  
the system of equations:
\BE
\label{2r}
\left( 
[(\frac{1}{2} \hat{J}''-{\omega^{(1)}_l}^2)\vec{V}_l+
(\hat{J}'-2\omega_l\omega^{(1)}_l)\vec{V}^{(1)}_l+
(\frac{1}{2}(\hat{J}-\omega^2_l) \vec{V}^{(2)})]
 \frac{\partial^2}{\partial s_1^2}+
2 i\omega_l \vec{V}_l \frac{\partial}{\partial t_2}\right)\,
A(s_1,t_2)+ \nonumber \\ 
\vec{\cal{Q}} |A(s_1,t_2)|^2 A(s_1,t_2) =0 
\EE
where 
\BE
\label{calq}
\vec{\cal{Q}}=
\sum_{d=0}^{2}\sum_{k=0}^{d} \sum_{\alpha'',\alpha'\leq \alpha''}
\vec{c}_{d,k}(\alpha',\alpha'')[ (i\omega_l)^k (-2i\omega_l)^{d-k}
V^*_l(\alpha')\gamma_l(\alpha'')+(-2i\omega_l)^k (i\omega_l)^{d-k}
V^*_l(\alpha'')\gamma_l(\alpha')]+ \nonumber\\
2\sum_{d=0}^{2}\sum_{\alpha'',\alpha'\leq \alpha''}
\vec{c}_{d,0}(\alpha',\alpha'')(-i\omega_l)^d \mu_l(\alpha')V_l(\alpha'')
+\nonumber \\
\sum_{d=0}^{2} \sum_{k=0}^{d}\sum_{j=0}^{k}
\sum_{\alpha''',\alpha''\leq \alpha''', \alpha'\leq \alpha'' } 
\vec{C}_{d,k,j}(\alpha',\alpha'',\alpha''')
[(-i\omega_l)^j(-i\omega_l)^{k-j} (i\omega_l)^{d-k}
  V_l(\alpha')V_l(\alpha'')V_l^*(\alpha''')+ \nonumber \\  
 (-i\omega_l)^j(i\omega_l)^{k-j} (-i\omega_l)^{d-k}
  V_l(\alpha')V^*_l(\alpha'')V_l(\alpha''')+
(i\omega_l)^j(-i\omega_l)^{k-j} (-i\omega_l)^{d-k}
  V^*_l(\alpha')V_l(\alpha'')V_l(\alpha''')] 
\EE

The first line of the equation (\ref{2r}) corresponds to the third
 order  expansion (\ref{o3}) of the linear  operator equation
 (\ref{d}) applied to the wave packet (terms in
 $e^{i(q_0n_0-\omega_l(q_0)t_0)}$ in (\ref{wpc})), in the moving
 reference frame used in (\ref{ls}),
 $<q_1>= -i ({\partial
 }/{\partial s_1}) A(s_1,t_2)$, and  $<\omega^{(2)}_l {q_1^2}/{2}> =
<\Delta(\Delta(\omega_l))>= i (\partial / \partial t_2 ) A(s_1,t_2)$.

The first two lines of  the  non linear coefficients vector
$\vec{\cal{Q}}$ arise from the double product, 
in the  nonlinear quadratic force terms in (\ref{emnl}),
between $O(\epsilon)$ and $O(\epsilon^2)$  components of (\ref{wpc}); 
the last two lines come from the nonlinear cubic
force terms in (\ref{emnl}) when considering just the
$O(\epsilon)$ terms in $\vec{E}_l(n,t)$.  
 
Multiplying Eq.~(\ref{2r}) by $\vec{V}_l^*$ we obtain the NLS equation:
\be
\label{nls}
\left( P \frac{\partial^2}{\partial s_1^2}+
i  \frac{\partial}{\partial t_2} \right) 
A(s_1,t_2)+  Q |A(s_1,t_2)|^2 A(s_1,t_2) =0 
\ee
where 
\be
\label{p}
P=\frac{1}{2\omega_l}
\left( \vec{V}^*_l \frac{\hat{J}''}{2}\vec{V}_l
- (\frac{ {\vec{V}_l}^* \hat{J}' \vec{V}_l}{2 \omega_l})^2
+ \sum_{m\neq l} \frac{| {\vec{V}_m}^* \hat{J}' \vec{V}_l|^2} 
{ \omega^2_l-\omega^2_m} \right)
\ee
and
\be
\label{q}
Q=\frac{\vec{V}^*_l \vec{\cal{Q}}}{2\omega_l}
\ee

The first part of equation (\ref{nls}) is  equation (\ref{ls}) for
the wave packet with $2P=\omega^{(2)}_l$ given by the (\ref{os}) to which
is now added the nonlinear part with coefficient $Q$.
If $PQ > 0$ then the effect of the amplitude dependent non linear
potential well in (\ref{nls}) balances the  wave packet group velocity
dispersion giving rise to the stable envelope soliton solution \cite{rem}. 

%%%%%%%%%%%%%%%%%%%%%%%%%%%%%%%%%%%%%%%%%%%%%%%%%%%%%%%%%%%%%%%%%%%%%%%
\section{Summary.}
The main outlines of the approach we introduce are  the following.
 The envelope soliton like solutions arise in systems
with weak dispersion and weak nonlinearity by two parallel 
 series expansions driven by a common expansion parameter ($\epsilon$):
on one hand the  weakness of the diffusion, for a wave packet like solution,
allows an expansion of the equations that
regulate the space time behaviour of the solution
on different scales; on the other hand the weak nonlinearity, for small
amplitude solutions, allows to write the equations of motion at
increasing orders of accuracy introducing the nonlinear terms in a
progressive way.
For scalar fields the Taylor series expansion of dispersion relations
gives directly the diffusive part for the envelope equations of
motion. Vectorial fields are instead characterized by a linear part
which gives rise, in the $q$ space, to 
 an eigenvalue (dispersion relations) and 
eigenvector (relative amplitude of the different components) 
problem (\ref{d}) ;
 to obtain the correct expansion in multiple scales it is then
necessary to apply the perturbation theory (\ref{v}),(\ref{omega}).
Finally, to combine this perturbative expansion with the 
nonlinear one, we antitransform the equations (\ref{o2}), (\ref{o3})
as done in (\ref{oe}), (\ref{2r}). 

Following the approach introduced in this paper it is straightforward
to derive the NLS equation for every non linear vectorial lattice
with on-site nonlinearities and with an arbitrary number of
components. For more complex systems this  could even be programmed in symbolic
languages to provide a fully automatic method.
 After having identified the nonlinear coefficients
$c_{d,k}^{\alpha}(\alpha',\alpha'')$ and  
$C_{d,k,j}^{\alpha}(\alpha',\alpha'',\alpha''')$ in the equation of
motion (\ref{emnl}), there are only algebraic systems to solve:
  one has to solve the eigenvectors $V_l(q_0)$ and
the eigenvalues $\omega_l(q_0)$ of the matrix $\hat{J}(q_0)$, 
then  the  systems (\ref{ga}) and  (\ref{mu}) for $\gamma_l$ and
$\mu_l$, and to derive
 $P$ from  (\ref{p}) and $Q$ from  
(\ref{calq}) and  (\ref{q}).
{From}  (\ref{nls}) one then obtains, if $PQ \geq 0$, the envelope
function $A(x_1,t_1,t_2)$ that, inserted into  (\ref{wpc})
together with the eigenvector correction  (\ref{vp}),  (\ref{am}), 
gives the complete $O(\epsilon^2)$ solution we are looking for.
\section*{Acknowledgments}

We  thank  R. Monasson for a critical reading of the manuscript.


\begin{thebibliography}{99}

\bibitem{rem}
M. Remoissenet, {\em Low Amplitude Breather and Envelope Solitons
in Quasi-one-dimensional Physical Models},
Physical Review B, {\bf 33}, N. 4, 2386 (1996);

\bibitem{opt}
A. Hasegawa, Y. Kodama, {\em Solitons in Optical Communications},
(Clarendon Press, Oxford, 1995); 


\bibitem{noi}
M. Barbi, S. Cocco, M. Peyrard, {\em Helicoidal Model for DNA
Openings}, submitted to Phys. Lett. A.

\bibitem{coh}
C. Cohen-Tannoudji, D. Diu, F. Lalo\"{e}, {\em M\'{e}canique
Quantique}, (Hermann, Paris, 1973).

\bibitem{nota}
Note that if $J(0)$ has some null columns the linear system solution 
is defined unless  that for some constant  components. They have to be added 
as order  $\epsilon$  terms in (\ref{wpc}), enter in the r.h.s. of (\ref{mu}), 
and are then solved with the $O(\epsilon^2)$ equations  \cite{noi}; the
final $O(\epsilon)$ result is in this case a combination of oscillating
envelope soliton and non oscillating soliton contributions.

\end{thebibliography}
\end{document}